# Quantum Lattice Kinetic Scheme for Solving Two-dimensional and Three-dimensional Incompressible Flows

Yang Xiao[1], Liming Yang[1, 2, 3, *], Chang Shu[4, †], Yinjie Du[1], Hao Dong[1], Jie Wu[1]

[1]Department of Aerodynamics, College of Aerospace Engineering, Nanjing University of Aeronautics and Astronautics, Nanjing 210016, China

[2]State Key Laboratory of Mechanics and Control for Aerospace Structures, Nanjing University of Aeronautics and Astronautics, Nanjing 210016, China

[3]MIIT Key Laboratory of Unsteady Aerodynamics and Flow Control, Nanjing University of Aeronautics and Astronautics, Nanjing 210016, China

[4]Department of Mechanical Engineering, National University of Singapore, Singapore 117576, Singapore

## Abstract

Lattice Boltzmann method (LBM) is particularly well-suited for implementation on quantum circuits owing to its simple algebraic operations and natural parallelism. However, most quantum LBMs fix $\tau = 1$ to avoid nonlinear collision, which restricts the simulation to a fixed mesh size for a given Reynolds number. To preserve the simplicity of setting $\tau = 1$ while enhancing flexibility, we propose a quantum lattice kinetic scheme (LKS) by introducing a constant parameter $A$ into the equilibrium distribution function (EDF), enabling independent adjustment of the fluid's viscosity.

─────────────────────

*Corresponding author, E-mail: lmyang@nuaa.edu.cn.

†Corresponding author, E-mail: mpeshuc@nus.edu.sg.

This modification removes the constraint on mesh size, making it possible to simulate flows with arbitrary Reynolds numbers. The Chapman-Enskog analysis confirms the modified EDF still recovers the Navier-Stokes equations without compromising collision accuracy. We evaluate the method on 2D and 3D Taylor-Green vortex and lid-driven cavity flows, demonstrating that quantum LKS attains the same accuracy and convergence order as classical LKS. The first application of quantum LBM to 3D incompressible flows represents a significant step forward in large-scale fluid dynamics simulation.

# 1 Introduction

In recent years, the unique features of quantum entanglement and superposition in quantum computing have enabled the development of quantum algorithms with applications across diverse fields such as finance [1-2], chemistry [3-4], machine learning [5-6], and optimization [7-8], garnering significant attention. Unlike classical bits in conventional computers, which are strictly limited to binary states 0 or 1, quantum bits (qubits) can exist in a superposition of both states simultaneously. Consequently, a quantum system with $n$ qubits can represent a linear combination of $2^n$ basis states, whereas a classical $n$-bit system can only represent a single one of these $2^n$ states [9]. These features offer the potential for quantum algorithms to outperform classical algorithms in efficiency and scalability.

Currently, the most widely recognized quantum algorithms that offer significant speedup over classical counterparts are Shor's algorithm for integer factorization [10] and Grover's algorithm for unstructured database search [11]. Essentially, these two algorithms are built upon quantum phase estimation (QPE) [12] and quantum amplitude amplification algorithms (QAAA) [13], respectively. These foundational algorithms have, in turn, inspired the development of quantum algorithms for the field of scientific computing. In computational fluid dynamics (CFD), for instance, the governing partial differential equations (PDEs) are typically discretized over a computational mesh to form a system of linear equations. This allows the problem of solving PDEs to be reduced to solving linear systems [14]. Leveraging this, Harrow, Hassidim, and Lloyd proposed the HHL algorithm based on QPE to solve linear systems with exponential speedup over classical algorithms [15]. This algorithm has since been integrated into CFD solvers as a quantum subroutine [16-17], demonstrating its potential to accelerate CFD simulations. Further advances include Kacewicz's [18] development of a quantum ordinary differential equation (ODE) solver using the quantum amplitude estimation algorithm (QAEA) [13], a QAAA-based method that achieves quadratic speedup compared to classical methods. Building on this, Gaitan [19] extended the approach to PDEs by first discretizing them into systems of ODEs, and Oz et al. [20] applied the method to simulate the Burgers' equation. Xiao et al. [9] further enhanced Gaitan's method by proposing a novel quantum Fourier ODE solver for simulating incompressible flows. These developments highlight increasing interest in quantum algorithms for CFD. Nevertheless, the development of quantum hardware is still in its early stages, with high error rates and circuit complexity that can scale exponentially [21], potentially negating any quantum advantage [22]. Despite this, quantum computing offers significant memory advantages due to its use of quantum superposition. For instance, storing flow variables with 64-bit precision across approximately $Re^{9/4} \approx 10^{18}$ mesh points would require around 8000 petabytes ($8 \times 10^9$ GB) on a classical computer, which is far beyond current computing capabilities [21, 23]. In contrast, by using amplitude encoding of classical information on quantum hardware, the required number of qubits is only about 60 (~15/2log(Re)), which is well within the capability of current quantum hardware [24].

In contrast to the macroscopic equation-based CFD algorithms, the lattice

Boltzmann method (LBM) operates on a mesoscopic level and has also garnered significant attention in CFD applications [25-26]. Owing to its simple algebraic operations and natural parallelism, LBM is particularly well-suited for implementation on quantum circuits. However, LBM demands additional storage for distribution functions, leading to higher memory requirements than macroscopic equation-based CFD algorithms [27]. Fortunately, this drawback is expected to be effectively alleviated by the memory advantages offered by quantum computing. This has spurred a wave of research into quantum LBM [28-33]. Mezzacapo et al. [28] developed the first quantum simulator based on the lattice kinetic formulation for fluid transport phenomena. Odorova and Steijl [29] introduced a quantum algorithm for the collisionless Boltzmann equation by implementing the streaming step via a quantum walk process. Building on the streaming step introduced in [29], Budinski [30] proposed the first quantum LBM for solving 1D and 2D linear advection-diffusion equations and later extended it to incompressible flows using the vorticity–streamfunction formulation [31]. Kocherla et al. [32] proposed a two-circuit quantum LBM that reduces quantum gate counts compared to single-circuit implementations, while Wawrzyniak et al. [33] applied quantum LBM to 3D advection-diffusion problems and proposed improved initialization and collision operations using (sublinear) amplitude encoding to enhance efficiency. However, a common limitation in [30 – 33] is the use of a fixed relaxation parameter $\tau = 1$ to eliminate the nonlinearity of the collision term, thereby reducing the right-hand side of the lattice Boltzmann equation to solely the equilibrium distribution function (EDF). This simplification fixes the viscosity to a constant value $\mu = 0.5\rho c_s^2 \delta t$, where $c_s$ is the lattice sound speed and $\delta t$ is the streaming time step, implying that the mesh size must also be fixed for a given viscosity. As a result, these methods are restricted to relatively simple, low-Reynolds-number flows. To address this issue, researchers such as Itan et al. [21, 34] and Sanavio et al. [35-36] applied Carleman linearization to eliminate the nonlinearity of the collision term. However, this approach leads to an exponential increase in the dimensionality of the linearized system [37–38].

In this work, we aim to develop a practical quantum LBM capable of efficiently handling nonlinear collision terms on relatively simple quantum circuits. Building on the lattice kinetic scheme (LKS) introduced by Inamuro [39], we develop a quantum LKS to simulate incompressible flows at arbitrary Reynolds numbers. Specifically, we retain the simplification $\tau = 1$ while introducing a constant parameter $A$ by modifying the EDF. Through Chapman-Enskog (C-E) expansion analysis, we establish a direct relationship between the viscosity $\mu$ and $A$, namely $\mu = \rho\left(0.5 - 2Ac_s^2\right)c_s^2\delta t$. This modification decouples the viscosity from the mesh size, thereby enabling simulations at varying Reynolds numbers on a fixed mesh size. The quantum LKS consists of four main steps: initialization, collision, streaming, and computation of macroscopic variables. The flow field is initialized in the quantum circuit using amplitude encoding. The collision step is implemented by decomposing the coefficient matrix into unitary operators via the linear combination of unitaries (LCU) method [40]. The streaming step is realized using a quantum walk, and macroscopic quantities are extracted through a sequence of SWAP and Hadamard gates. Since $\tau = 1$ is preserved, the proposed quantum LKS can be directly integrated into frameworks [30 – 33], enhancing its flexibility and extensibility. To assess the effectiveness of the proposed method, we first analyze its convergence order and then validate its accuracy using several representative 2D and 3D incompressible flows. To the best of our knowledge, this is the first work to apply quantum LBM to 3D incompressible flow simulations.

## 2 Results

### 2.1 Quantum lattice kinetic scheme

In this section, we will detail the implementation of the quantum LKS. As shown in Fig. 1(a), the quantum LKS consists of four main components: initialization, collision, streaming, and computation of macroscopic variables. The quantum circuit of the quantum LKS comprises five registers. The qubit registers $\left|q_x\right\rangle$, $\left|q_y\right\rangle$, and $\left|q_z\right\rangle$ denote spatial dimensions embedding, each requiring $n_d = \log_2(\{M_x, M_y, M_z\})$ qubits, where the subscript $d = \{x, y, z\}$ and $M_x, M_y, M_z$ are the number of lattice sites in each spatial direction. The qubit register $\left|q_Q\right\rangle$ encodes the distribution functions, requiring $n_Q = \text{ceil}(\log_2 Q)$ qubits, where $Q$ denotes the number of discrete velocity directions in the LBM. The ancillary register $\left|a_0\right\rangle$ is used to support the implementation of the collision and macroscopic variable computation blocks, with a fixed number of $n_a = 1$ qubit. Consequently, the total number of qubits required for the quantum LKS circuit is given by $n_{total} = \sum_{i \in d} n_i + n_Q + n_a$.

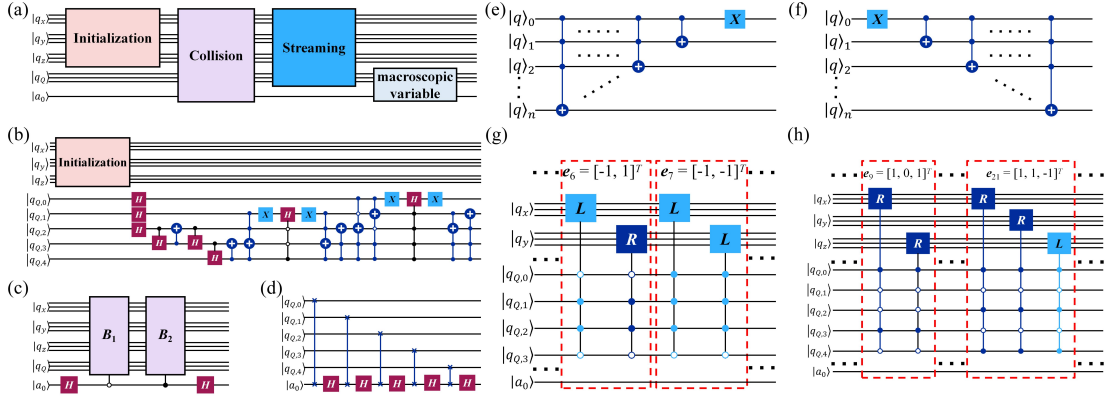

**Fig. 1 Schematic diagram of the proposed quantum LKS.** (a) Schematic of the quantum circuit with general blocks for quantum LKS. (b) Quantum circuits for the initialization process of the D3Q27 lattice model. (c) Quantum circuit of the collision operator. (d) Quantum circuits for computing the macroscopic density in D3Q27 lattice model. (e) Quantum circuits for the streaming operator $\boldsymbol{R}$. (f) Quantum circuits for the streaming operator $\boldsymbol{L}$. (g) Quantum circuits for the streaming step in selected directions $\boldsymbol{e}_6 = [-1, 1]^T$ and $\boldsymbol{e}_7 = [-1, -1]^T$ for the D2Q9 lattice model. (h) Quantum circuits for the streaming step in selected directions $\boldsymbol{e}_9 = [-1, 0, 1]^T$ and $\boldsymbol{e}_{21} = [1, 1, -1]^T$ for the D3Q27 lattice model.

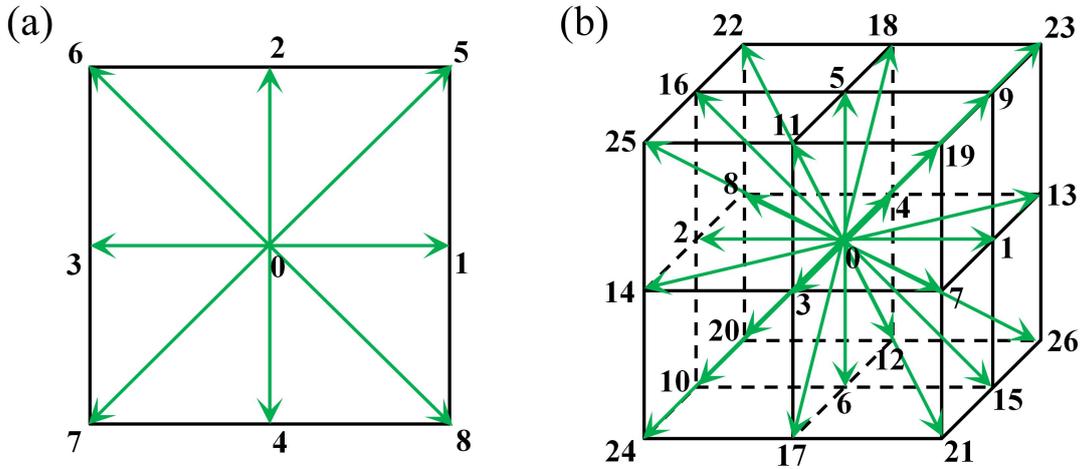

**Fig. 2** Structures of the lattice unit cell for (a) D2Q9 and (b) D3Q27 models.

In this work, we employ the D2Q9 and D3Q27 lattice models to simulate 2D and 3D incompressible flows, respectively. The structures of the D2Q9 and D3Q27 lattice models are illustrated in Fig. 2. Notably, when simulating 2D flows, the qubit register $|q_z\rangle$ can be omitted without loss of generality.

### 2.1.1 Initialization step with duplication sequence

In the initialization step, the macroscopic density field is encoded into the quantum circuit via amplitude encoding. This procedure follows the method proposed by Shende et al. [42], which was later shown by Wawrzyniak et al. [33] to require fewer quantum gates for encoding the density field, thereby reducing the computational cost in classical simulations of quantum circuits. First, the initial density field $\boldsymbol{\rho}_{init}$ is reshaped into a column vector $\boldsymbol{\rho}$ of size $M = M_x \cdot M_y \cdot M_z$. The reshaped vector $\boldsymbol{\rho}$ is then encoded into the spatial registers $|q_d\rangle$, yielding the following initialized quantum state $|\psi_A\rangle = |0\rangle_a |000\rangle_Q \frac{1}{\|\boldsymbol{\rho}\|} \sum_{k=0}^{2^{N_{init}}-1} \rho_k |k\rangle$, where $N_{init} = \sum_{i \in d} n_i$, and $\|\bullet\|$ is the Euclidean norm.

Next, the Hadamard and controlled-Hadamard gates are applied to the velocity-direction register $|q_Q\rangle$ to duplicate the initialized density field across its qubit subspace, preparing it for further processing in subsequent steps. The quantum circuit used for this initialization process in the D3Q27 lattice model is shown in Fig. 1(b), while the corresponding circuit for the D2Q9 lattice model is provided in Fig. S1 of the Supplementary Information. Taking D2Q9 as an example, the Hadamard gates are applied to the first two qubits of the $\left|q_Q\right\rangle$ register, resulting in the evolution of the state to: $\left|\psi_A\right\rangle = \left|0\right\rangle_a \left( \frac{1}{2}\left|0000\right\rangle_Q + \frac{1}{2}\left|0001\right\rangle_Q + \frac{1}{2}\left|0010\right\rangle_Q + \frac{1}{2}\left|0011\right\rangle_Q \right) \frac{1}{\|\boldsymbol{\rho}\|} \sum_{k=0}^{2^{N_{init}}-1} \rho_k \left|k\right\rangle.$

By applying subsequent controlled gates, the quantum state further evolves to:

$$\begin{aligned}
\left|\psi_B\right\rangle = &\left|0\right\rangle_a \left(\frac{1}{2}\left|0000\right\rangle_Q + \frac{1}{2\sqrt{2}}\left|0001\right\rangle_Q + \frac{1}{2\sqrt{2}}\left|0010\right\rangle_Q + \frac{1}{2\sqrt{2}}\left|0011\right\rangle_Q \right. \\
&\left. + \frac{1}{4}\left|0100\right\rangle_Q + \frac{1}{4}\left|0101\right\rangle_Q + \frac{1}{4}\left|0110\right\rangle_Q + \frac{1}{4}\left|0111\right\rangle_Q + \frac{1}{2\sqrt{2}}\left|1000\right\rangle_Q \right) \frac{1}{\|\boldsymbol{\rho}\|} \sum_{k=0}^{2^{N_{init}}-1} \rho_k \left|k\right\rangle.
\end{aligned} \quad (1)$$

As shown in Eq. (1), each subspace of the register $\left|q_Q\right\rangle$ corresponds to the distribution function in a specific velocity direction. For the D3Q27 lattice model, the coefficients of the subspaces in the register $\left|q_Q\right\rangle$ can be derived similarly and are given by the sequence [$1/2\sqrt{2}$ , $1/2\sqrt{2}$ , $1/2\sqrt{2}$ , $1/2\sqrt{2}$ , $1/4\sqrt{2}$ , $1/4\sqrt{2}$ , $1/4\sqrt{2}$ ,

$1/4\sqrt{2}$ , $1/4\sqrt{2}$ , $1/4\sqrt{2}$ , $1/4\sqrt{2}$ , $1/4\sqrt{2}$ , 1/8, 1/8, 1/8, 1/8, $1/4\sqrt{2}$ , $1/4\sqrt{2}$ ,

$1/4\sqrt{2}$ , $1/4\sqrt{2}$ , 1/8, 1/8, $1/8\sqrt{2}$ , $1/8\sqrt{2}$ , 1/16, 1/16, $1/8\sqrt{2}$ ].

### 2.1.2 Collision step based on the LCU method

Since the operations of quantum computing are inherently linear, we similarly set

the relaxation parameter $\tau = 1$ and adopt the EDF defined in Eq. (19). This choice

simplifies the LBE to Eq. (15) and enables the simulation of arbitrary Reynolds

numbers without imposing constraints on the mesh size. Consequently, the entire

computational process of the quantum LKS can be expressed as $Uf_\alpha^n = f_\alpha^{n+1}$, where

$f_\alpha^n$ and $f_\alpha^{n+1}$ represent the distribution functions at time steps $n$ and $n + 1$,

respectively. The operator $U$ is then decomposed into a series of unitary operators and encoded into the quantum circuit. According to the initialization procedure described in Section 4 and the EDF defined in Eq. (19), the coefficient matrix used to implement the collision step in the quantum LKS is given by

$$D = \begin{bmatrix} C_0 D_0 & 0 & 0 \\ 0 & \ddots & 0 \\ 0 & 0 & C_{Q-1} D_{Q-1} \end{bmatrix}, \qquad D_\alpha = \begin{bmatrix} w_\alpha \hat{f}_\alpha^{eq}(\boldsymbol{x}_0, t) & 0 & 0 \\ 0 & \ddots & 0 \\ 0 & 0 & w_\alpha \hat{f}_\alpha^{eq}(\boldsymbol{x}_{M-1}, t) \end{bmatrix}, \qquad (2)$$

where

$$\hat{f}_\alpha^{eq}(\boldsymbol{x}, t) = f_\alpha^{eq}(\boldsymbol{x}, t)/\rho = w_\alpha \left( 1 + \frac{e_\alpha \cdot \boldsymbol{u}}{c_s^2} + \frac{(e_\alpha \cdot \boldsymbol{u})^2}{2c_s^4} - \frac{\boldsymbol{u} \cdot \boldsymbol{u}}{2c_s^2} + A \partial e_\alpha^T \left( \nabla \boldsymbol{u} + (\nabla \boldsymbol{u})^T \right) e_\alpha \right). \qquad (3)$$

Here, $C_\alpha = \sqrt{2}^h$ is a constant determined by the replication sequence during initialization, where $h$ represents the number of Hadamard gates applied to the scalar field. To perform the collision step, the coefficient matrix $D$ must be applied to the initialized state $|\psi_B\rangle$. However, since $D$ is a non-unitary matrix, it is reconstructed as a linear combination of unitary matrices using the LCU method [40], i.e., $D = (B_1 + B_2)/2$, where $B_1$ and $B_2$ are the unitary matrices defined as $B_1 = D + i\sqrt{I - D^2}$ and $B_2 = D - i\sqrt{I - D^2}$, and $I$ denotes the identity matrix.

As illustrated in Fig. 1(c), the collision operator can be represented as

$$\left( H^* \otimes I^{\otimes n_{total}-1} \right) \left( |0\rangle\langle 0|_a \otimes B_1 + |1\rangle\langle 1|_a \otimes B_2 \right) \left( H \otimes I^{\otimes n_{total}-1} \right) \qquad (4)$$

Applying the operator in Eq. (4) to the initialized state in Eq. (1) yields

$$\begin{aligned} |\psi_C\rangle &= |0\rangle_a D|\psi_B\rangle + |1\rangle_a \left( i\sqrt{I - D^2} \right) |\psi_B\rangle \\ &= |0\rangle_a \frac{1}{\|\boldsymbol{\rho}\|} \sum_{\alpha=0}^{Q-1} \sum_{k=0}^{2^{N_{init}}-1} f_\alpha^{eq}(\boldsymbol{x}_k, t) |\alpha\rangle_Q |k\rangle + |1\rangle_a \left( i\sqrt{I - D^2} \right) |\psi_B\rangle. \end{aligned} \qquad (5)$$

It can be seen that the result of the collision step is successfully realized when the auxiliary register $|a_0\rangle$ is measured in the $|0\rangle$ state. Therefore, the operator $D$ is effectively applied only upon successful measurement of the ancilla in state $|0\rangle$; otherwise, the operation must be repeated.

### 2.1.3 Streaming step based on the quantum walk

In the streaming step, the distribution function $f_\alpha$ propagates along its corresponding particle velocity direction $\boldsymbol{e}_\alpha$, which can be realized using quantum walk operations [29-30]. Specifically, two unitary operators $\boldsymbol{R}$ and $\boldsymbol{L}$ are introduced, whose matrix forms are given by

$$\boldsymbol{R} = \begin{bmatrix} 0 & 0 & 0 & \cdots & 0 & 1 \\ 1 & 0 & 0 & \cdots & 0 & 0 \\ 0 & 1 & 0 & \cdots & 0 & 0 \\ \vdots & \vdots & \vdots & \ddots & \vdots & \vdots \\ 0 & 0 & 0 & \cdots & 0 & 0 \\ 0 & 0 & 0 & \cdots & 1 & 0 \end{bmatrix}, \qquad \boldsymbol{L} = \begin{bmatrix} 0 & 1 & 0 & \cdots & 0 & 0 \\ 0 & 0 & 1 & \cdots & 0 & 0 \\ 0 & 0 & 0 & \cdots & 0 & 0 \\ \vdots & \vdots & \vdots & \ddots & \vdots & \vdots \\ 0 & 0 & 0 & \cdots & 0 & 1 \\ 1 & 0 & 0 & \cdots & 0 & 0 \end{bmatrix}, \qquad (6)$$

where the operator $\boldsymbol{R}$ corresponds to a right (or positive) shift operation, whereas the operator $\boldsymbol{L}$ corresponds to a left (or negative) shift operation. The quantum circuits of operators $\boldsymbol{R}$ and $\boldsymbol{L}$ are composed of multi-controlled X-gates, as shown in Figs. 1(e)(f).

The streaming operations are applied solely to the register $\left| q_d \right\rangle$, and conditionally directed toward specific lattice links by imposing control conditions on the register $\left| q_Q \right\rangle$. Figs. 1(g)(h) present the quantum circuits for the streaming operations in selected particle velocity directions for the D2Q9 and D3Q27 lattice models. By applying the streaming operator corresponding to the particle velocity direction to the state $\left| \psi_C \right\rangle$, the system evolves into the following state

$$\left| \psi_D \right\rangle = \left| 0 \right\rangle_a \frac{1}{\|\boldsymbol{\rho}\|} \sum_{\alpha=0}^{Q-1} \sum_{k=0}^{2^{N_{init}}-1} f_\alpha^{eq}\left(\boldsymbol{x}_k - e_\alpha \delta t, t\right) \left| \alpha \right\rangle_Q \left| k \right\rangle. \qquad (7)$$

As with the collision step, the result of the streaming operation is observed in the auxiliary register $|a_0\rangle$ when it is in state $|0\rangle$. Notably, the streaming operator inherently satisfies periodic boundary conditions, thereby eliminating the need for additional boundary treatment in flows with periodic boundaries.

**2.1.4 Computation of macroscopic variables**

According to Eq. (17), the macroscopic density is computed by summing over all distribution functions. In the quantum circuit, this corresponds to summing the distribution functions encoded in the various subspaces of the register $|q_Q\rangle$, which can be accomplished by applying a series of SWAP and Hadamard gates to the registers $|q_Q\rangle$ and $|a_0\rangle$. The SWAP gates serve to rearrange the distribution functions, while the Hadamard gates are used to sum the state amplitudes. Fig. 1(d) presents the quantum circuits used for computing the macroscopic density in the D3Q27 lattice model, while the D2Q9 lattice model can be found in Fig. S2 of the Supplementary Information. It can be observed that the complexity of the quantum circuit depends solely on the number of qubits in the register $|q_Q\rangle$, meaning that it scales with the size of the velocity set rather than the mesh size. By applying the quantum circuits in Fig. 1(d) to the state $|\psi_D\rangle$ and then measuring the auxiliary register $|a\rangle$ in the $|0\rangle$ state yields $\boldsymbol{\rho}^* = 1 / \left( \sqrt{2}^{n_Q} \|\boldsymbol{\rho}\| \right) \sum_{\alpha=0}^{Q-1} f_\alpha^{eq} (\boldsymbol{x} - e_\alpha \delta t, t)$. Comparing this with Eq. (17), it is clear that the macroscopic density $\rho(\boldsymbol{x}, t + \delta t)$ at the next time step can be recovered by multiplying the output $\boldsymbol{\rho}^*$ of the quantum circuit by a constant $\sqrt{2}^{n_Q} \|\boldsymbol{\rho}\|$.

However, quantum circuits in their current form are limited to computing the 0th moment of the distribution function. To compute the momentum term $\rho\boldsymbol{u}$, which corresponds to the 1st moment, we follow the two-circuit quantum LBM proposed by Kocherla et al. [32]. Specifically, we employ multiple quantum circuits with identical architecture to separately compute macroscopic density and momentum. The complete computational framework of the quantum LKS for simulating 3D incompressible flows is illustrated in Fig. 3. In this framework, the initial density field $\rho$ serves as the input for computing the macroscopic density, while the product of the initial density and the particle velocity $\boldsymbol{e}_\alpha \rho$ serves as the input for computing the macroscopic momentum $\rho \boldsymbol{u}$. This allows both quantities to be extracted from the 0th moment via parallel quantum circuits.

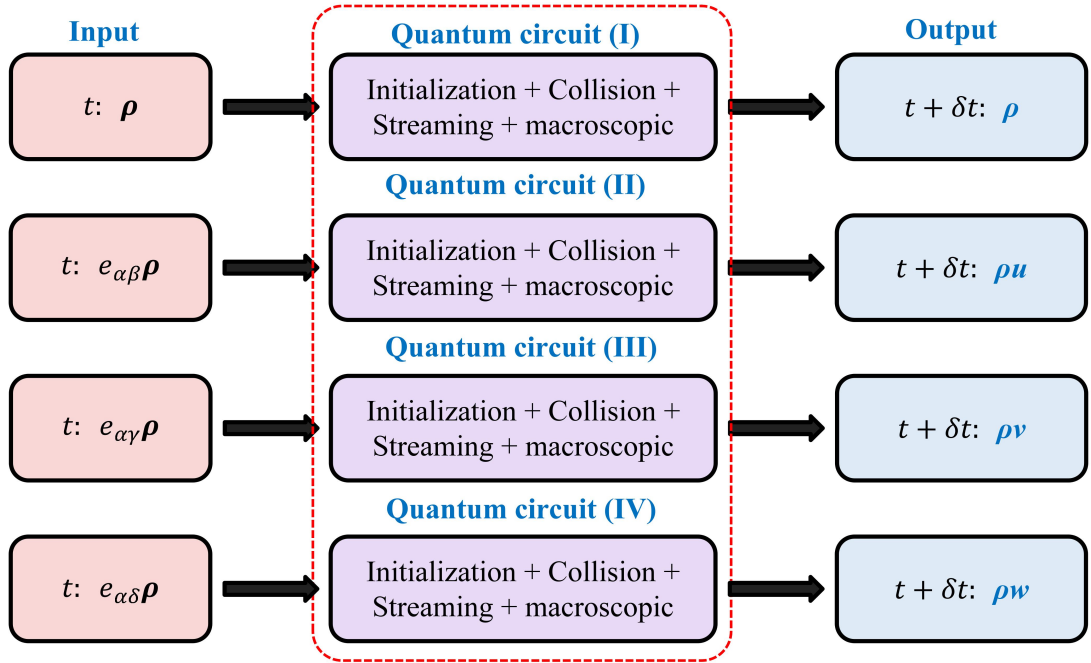

**Fig. 3** Schematic of the quantum LKS for simulating incompressible flows.

## 2.1.5 Computational complexity

In this section, we analyze the computational complexity of the proposed quantum LKS algorithm, which consists of four components: initialization, collision, streaming, and computation of macroscopic quantities. Since CNOT gates are significantly slower to execute than single-qubit gates [32], our analysis primarily focuses on the number of CNOT gates involved in the initialization and collision steps. For the streaming and macroscopic computation stages, the complexity analysis emphasizes the use of multi-controlled X and SWAP gates, which dominate the operations in those components.

The initialization step involves encoding the input data. In Qiskit, data encoding follows the isometric synthesis method described by Iten et al. [43], which requires $O(2^{n_x+n_y+n_z})$ CNOT gates to encode arbitrary data of dimension $2^{n_x+n_y+n_z}$, where $n_x = \log_2 M_x$, $n_y = \log_2 M_y$, and $n_z = \log_2 M_z$. In comparison with the algorithms in [30–33], which encode the initial density field separately for each discrete velocity direction with a total complexity of $O(Q \cdot 2^{n_x+n_y+n_z})$, our selected method encodes the initial density field only once. This reduces the exponential computational complexity by a factor of $Q$, thereby significantly enhancing computational efficiency.

The collision step is implemented using diagonal gates in Qiskit. According to Shende et al. [44], constructing a diagonal gate for a $2^n \times 2^n$ dimensional diagonal matrix requires $O(2^n)$ CNOT gates. Since the LCU Method yields two such diagonal operators for the collision step in each velocity direction, the total computational complexity is $O(Q \cdot 2^{n_x+n_y+n_z+1})$.

For the streaming step, as shown in Figs. 1(e)(f), the **R** and **L** operators are applied along the directions with non-zero particle velocities. These operators are composed of multi-controlled X gates, with the number of such gates being proportional to the number of spatial mesh points in each direction. Assuming uniform discretization in all spatial directions, i.e., $M_x = M_y = M_z$, the computational complexity of the streaming step is $O(\sigma \cdot n_x)$, where $\sigma$ denotes the number of non-zero velocity directions and $n_x = \log_2 M_x$.

The computation of macroscopic quantities involves only SWAP and Hadamard gates. This step is independent of the number of spatial grid points and depends solely on the size of the velocity set, resulting in a computational complexity of $O(2 \cdot n_Q)$, where $n_Q = \text{ceil}(\log_2 Q)$.

In summary, the overall computational complexity of the quantum LKS algorithm is $O[\left(2Q+1\right) \cdot 2^{n_x+n_y+n_z} + \sigma \cdot n_x + 2 \cdot n_Q]$.

## 2.2 Numerical Examples

### 2.1.1 Two-dimensional incompressible flows

In this section, we apply the proposed quantum LKS based on the D2Q9 lattice model to solve 2D incompressible flows at various Reynolds numbers. All simulations are carried out on a PC with an Intel i7-14700KF CPU. The implementation utilizes the Qiskit SDK version 0.45.2 and the AER backend version 0.13.2 [45].

### 2.2.1.1 Two-dimensional Taylor-Green vortex flow

In this section, we investigate the performance of the quantum LKS by solving the 2D Taylor-Green vortex flow, whose analytical solution can be found in Eq.

(SI-2.1) of the Supplementary Information. The computational domain is specified as $[-L, L] \times [-L, L]$, and the Reynolds number is defined as $Re = u_0 L / \upsilon$. All boundaries are set to periodic boundary conditions, and the initial conditions are provided by the analytical solution. The parameters used are: $\delta_t = 1$, $\Delta x = 1$, $\rho_0 = 1$, $\upsilon = 0.08$, and $Re = 10$. To validate the quantum LKS, we first compare the contours of the dimensionless horizontal velocity $u/u_0$ and vertical velocity $v/u_0$ at the dimensionless time $t^* = u_0 t / L = 1.0$ with results from the classic LKS and the analytical solutions. A uniform mesh size of $64 \times 64$ is used in this case. As illustrated in Figs. 4(a)(b), the velocity contours obtained from the quantum LKS exhibit excellent agreement with both the classic LKS results and the analytical solution.

To further assess the accuracy, we analyze the convergence order of the quantum LKS. Simulations are conducted on four different uniform meshes ($N \times N$, $N = 8$, 16, 32, and 64), with the characteristic velocity $u_0$ set to 0.2, 0.1, 0.05, and 0.025, respectively. The solutions of $u$-velocity are extracted at the dimensionless time $t^* = u_0 L / t = 1.0$, and the relative error is evaluated using the $L_2$ norm defined as $L_2 \text{ error} = \sqrt{\sum \left[ (u_{\text{numerical}} - u_{\text{exact}}) / u_0 \right]^2 / N_{\text{total}}}$, where $u_{\text{numerical}}$, $u_{\text{exact}}$, and $N_{\text{total}}$ are the numerical solution, analytical solution, and the total number of mesh points, respectively. As shown in Fig.4(c), both the quantum LKS and classic LKS achieve a convergence order close to 2, aligning with the theoretical second-order accuracy of the LBM. Table 1 summarizes the transient numerical errors of quantum LKS and classic LKS for different mesh sizes. The results indicate that the transient numerical errors of both methods are nearly identical, confirming the accuracy and validity of the quantum LKS.

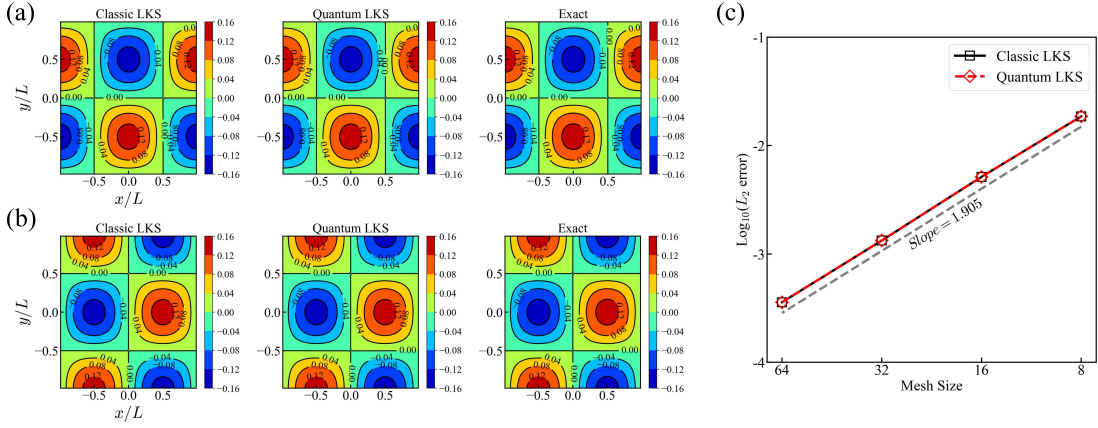

**Fig. 4 Two-dimensional Taylor-Green vortex flow.** Comparison of (a) *u*- and (b) *v*-velocities for the 2D Taylor-Green vortex flow obtained by classic LKS and quantum LKS with the mesh size of 64 × 64, and analytic solution. (c) Convergence order of the $L_2$ error versus mesh size.

**Table 1** $L_2$ error of the *u*- and *v*-velocities for the 2D Taylor-Green vortex flow.

| Mesh size | $u_0$ | $u/u_0$: $L_2$ error | | $v/u_0$: $L_2$ error | |
|:---:|:---:|:---:|:---:|:---:|:---:|
| | | Classic LKS | Quantum LKS | Classic LKS | Quantum LKS |
| 8×8 | 0.2 | 1.861E-2 | 1.862E-2 | 1.861E-2 | 1.862E-2 |
| 16×16 | 0.1 | 5.143E-3 | 5.146E-3 | 5.143E-3 | 5.146E-3 |
| 32×32 | 0.05 | 1.330E-3 | 1.330E-3 | 1.330E-3 | 1.330E-3 |
| 64×64 | 0.025 | 3.583E-4 | 3.583E-4 | 3.583E-4 | 3.584E-4 |

**2.2.1.2 Two-dimensional lid-driven cavity flow**

To further verify the robustness of the quantum LKS in a more practical problem,

we investigate the 2D lid-driven cavity flow, a classic benchmark test for newly developed numerical methods. The velocity on the walls follows the no-slip condition (Dirichlet boundary condition). All walls remain stationary except for the top wall, which moves at a constant velocity of $u = 0.1$. Unlike the periodic boundary conditions used in the Taylor-Green vortex flow in Section 2.2.1.1, the Dirichlet boundary conditions here are enforced on a classical computer after the quantum circuit has completed the computation of macroscopic quantities.

Simulations are performed on three different uniform meshes ($N \times N$, $N = 16$, 32, and 64), and three different Reynolds numbers of 100, 400, and 1000 are considered. The parameters are set as: $\delta_t = 1$, $\Delta x = 1$, and $\rho_0 = 1$. Figs. 5(a)(b) presents a comparison of the velocity profiles along the vertical and horizontal centerlines obtained from quantum LKS, classic LKS, and the benchmark results by Ghia et al. [46]. Clearly, the numerical results of quantum LKS are essentially consistent with those of classic LKS at the same mesh size, and both methods exhibit improved accuracy with increasing grid resolution. In particular, both quantum LKS and classic LKS produce reasonable results for $Re = 100$ even with a coarse mesh. While for $Re \geqslant 400$, a mesh size of at least $64 \times 64$ is required for both methods to achieve relatively accurate results. This can be more intuitively observed in the flow pattern shown in Figs. 5(c)(e)(g), where only at a mesh size of $64 \times 64$ can quantum LKS successfully capture the two secondary vortices at the bottom. This behavior arises from the coarse meshes introducing significant truncation errors, which degrade accuracy. Finally, Figs. 5(d)(f)(h) shows the convergence histories of both quantum

LKS and classic LKS across various mesh sizes and Reynolds numbers. As observed, the convergence histories of both methods align closely in all scenarios. This case further validates the accuracy and reliability of the quantum LKS.

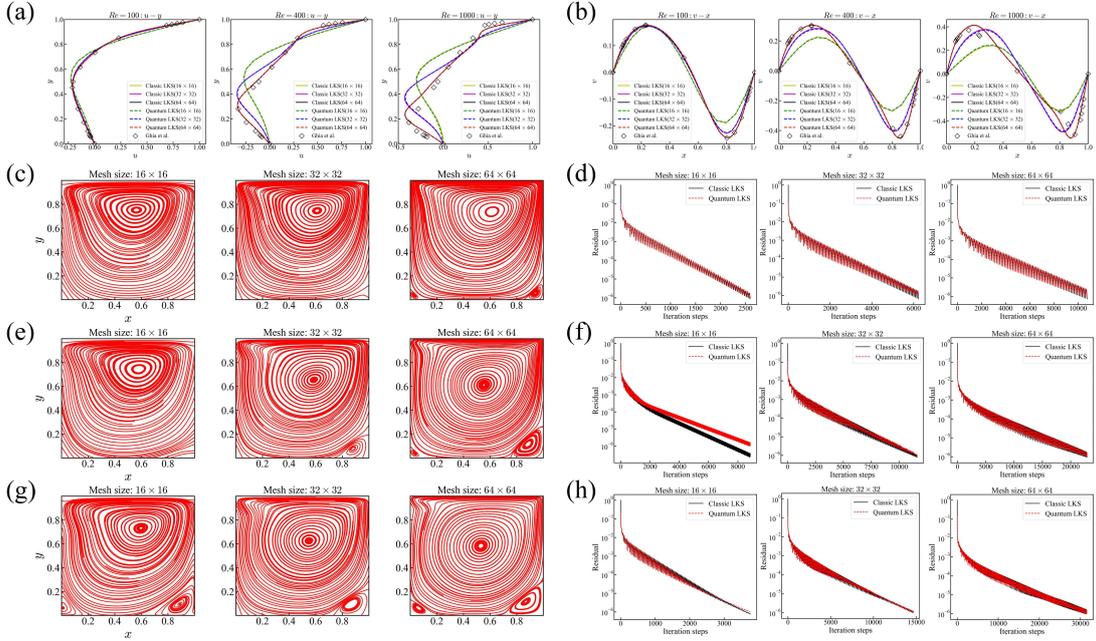

**Fig. 5 Two-dimensional lid-driven cavity flow.** Comparison of (a) vertical and (b) horizontal centerline velocities for the 2D lid-drive cavity flow with the results of Ghia et al. [46]. Streamlines of the 2D lid-driven cavity flow at (c) *Re* = 100, (e) *Re* = 400, and (g) *Re* = 1000 obtained by quantum LKS with different mesh sizes. Convergence history of quantum and classic LKS with different mesh sizes for the 2D lid-driven cavity flow at (d) *Re* = 100, (f) *Re* = 400, and (h) *Re* = 1000.

### 2.2.2 Three-dimensional incompressible flows

In this section, we apply the proposed quantum LKS based on the D3Q27 lattice model to simulate 3D incompressible flows.

### 2.2.2.1 Three-dimensional Taylor-Green vortex flow

In this case, we simulate the 3D Taylor-Green vortex flow. Specifically, we assume no flow in the $Z$ direction, and its analytical solution can be found in Eq. (SI-2.2) of Supplementary Information. The computational domain is specified as $[-L, L] \times [-L, L] \times [-L, L]$, and the Reynolds number is $Re = u_0 L / v = 100$. Similar to Section 2.2.1.1, periodic boundary conditions are applied on all sides, and the initial conditions are specified by the analytical solution. The parameters are set as: $\delta_t = 1$, $\Delta x = 1$, $\rho_0 = 1$, and $u_0 = 0.02$.

Simulations are conducted using a uniform mesh with the size of $N \times N \times N = 64 \times 64 \times 64$, and the computation is carried out up to the dimensionless time $t^* = u_0 t / L = 0.2$. Table 2 summarizes the transient numerical errors of horizontal velocity $u/u_0$ and vertical velocity $v/u_0$ obtained by quantum LKS and classic LKS. It can be observed that the transient relative errors of both methods exhibit a high degree of consistency. Fig. 6 compares the contour plots of the velocity components obtained by quantum LKS and classic LKS with the analytical solution. The close agreement further demonstrates the accuracy of the quantum LKS.

**Table 2** $L_2$ error of the $u$- and $v$-velocities for the 3D Taylor-Green vortex flow.

| Mesh size | $u_0$ | $u/u_0$: $L_2$ error | | $v/u_0$: $L_2$ error | |
|---|---|---|---|---|---|
| | | Classic LKS | Quantum LKS | Classic LKS | Quantum LKS |
| 64×64×64 | 0.02 | 6.838E-3 | 6.811E-3 | 6.838E-3 | 6.811E-3 |

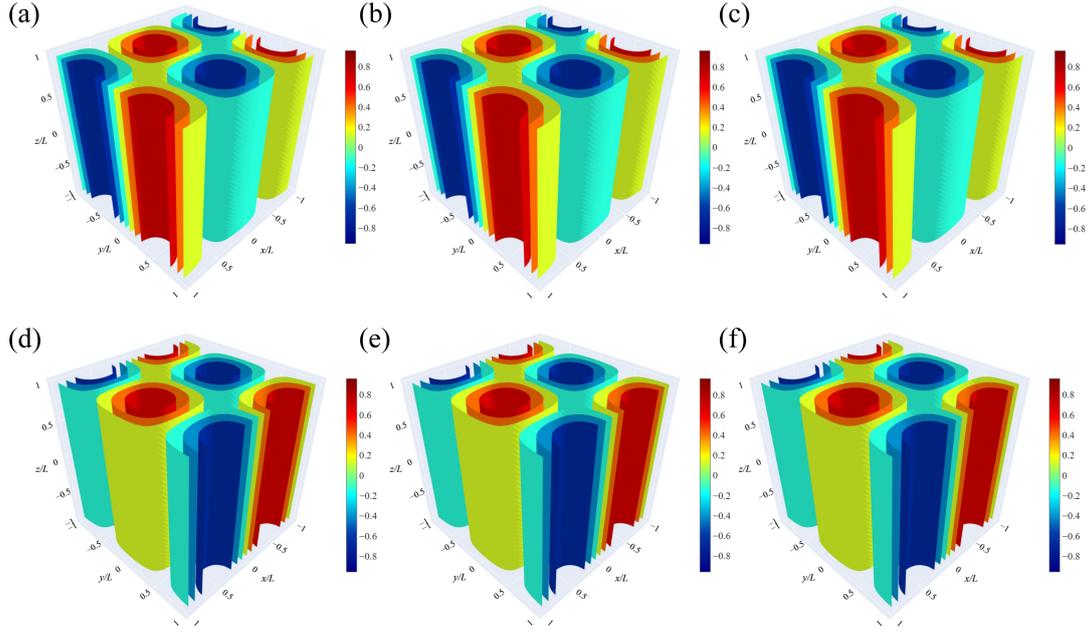

**Fig. 6 Three-dimensional Taylor-Green vortex flow.** Comparison of $u$- (Top) and $v$- (Bottom) velocities for the 3D Taylor-Green vortex flow obtained by (a) (d) classic LKS, (b) (e) quantum LKS, and (c) (f) analytic solution.

### 2.2.2.2 Three-dimensional lid-driven cavity flow

In this final test case, we simulate the more complex 3D lid-driven cavity flow. The velocity on all walls follows the no-slip condition (Dirichlet boundary condition), i.e., $u = 0$, $v = 0$, and $w = 0$ are applied to all five stationary walls, while $u = u_0$, $v = 0$, and $w = 0$ are applied to the top lid. The parameters are set as: $\delta_t = 1$, $\Delta x = 1$, $\rho_0 = 1$, $u_0 = 0.1$, and $Re = u_0 L/v = 100$. In the simulation, a uniform mesh with the size of $32 \times 32 \times 32$ is used. Figs. 7(a)(b) presents the velocity vector fields on the central $x$-$z$, $y$-$z$, and $x$-$y$ planes of the cavity, obtained by both the quantum LKS and the classic LKS. The two methods demonstrate strong agreement in the overall flow structure. Additionally, Fig. 7(c) shows the distribution of the $u$-velocity component along the vertical centerline of the cavity. The results align closely with those reported by Wong and Baker [47] and Jiang et al. [48], demonstrating excellent agreement. Finally, Fig. 7(d) presents the convergence history of the quantum LKS and classic LKS. The close alignment between the two confirms the stability and accuracy of the quantum LKS. This test case further validates the effectiveness and robustness of the quantum LKS in simulating complex 3D incompressible flows.

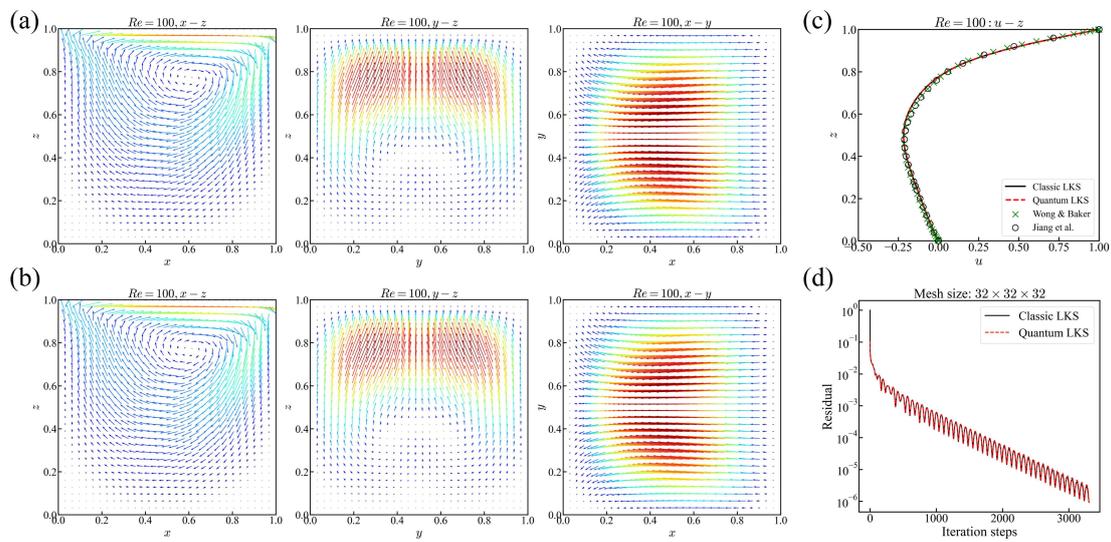

**Fig. 7 Three-dimensional lid-driven cavity flow.** Comparison of the flow patterns in vector form on the *x-z* centroidal plane, *y-z* centroidal plane, and *x-y* centroidal plane obtained by (a) classic LKS and (b) quantum LKS with the mesh size of 32 × 32× 32. (c) Comparison of the *u*-velocity component distribution along the vertical centerline with results from Wong and Baker [47] and Jiang et al. [48]. (d) Convergence history of quantum and classic LKS for the 3D lid-driven cavity flow at *Re* = 100.

## 3 Discussion

In this work, we proposed a quantum LKS for simulating 2D and 3D incompressible flows at arbitrary Reynolds numbers. Within the quantum LKS framework, the relaxation parameter is set to 1, reducing the right-hand side of the LBE to the EDF alone. This simplification enables the collision and streaming steps to be conveniently implemented via a sequence of unitary quantum operators. To enable flexible viscosity adjustment, we introduce a constant parameter $A$ into the EDF, as suggested by Inamuro [39]. The quantum LKS represents a general and versatile algorithm that alleviates the mesh size constraints encountered in previous quantum LBM while remaining fully compatible with their frameworks. This makes it straightforward to integrate and extend. Moreover, the initialization technique employed significantly reduces the required quantum gate count, thereby enhancing the overall computational efficiency.

The performance of the proposed method is evaluated via a series of benchmark tests involving representative 2D and 3D incompressible flows. Numerical results indicate that the solutions obtained by the quantum LKS closely match the classic LKS under the same mesh size, with both showing consistent error behavior relative to the analytical solution. Moreover, the convergence histories of the quantum and classic LKS are remarkably consistent across all test cases. In particular, both the quantum and classic LKS exhibit the same convergence order, consistent with the theoretical expectations of the LBM. These results confirm the adaptability and accuracy of the quantum LKS in simulating 2D and 3D incompressible flows.

In summary, the quantum LKS offers an effective pathway for simulating incompressible flows on quantum hardware. With the memory advantages of quantum computing, it holds a great potential for advancing large-scale fluid dynamics simulations in future engineering applications. However, the quantum LKS remains essentially a hybrid quantum-classic algorithm, as the computation of gradient terms in the EDF and the treatment of non-periodic boundary conditions still rely on classical computation. Future research will focus on developing a fully quantum version of the quantum LKS. Furthermore, since the current approach requires separate executions of identical quantum circuits to compute macroscopic density and velocity, an important direction will be to explore strategies for extracting both quantities within a single quantum circuit.

## 4 Methods

The single relaxation time lattice Boltzmann equation (LBE) is formulated as

$$f_\alpha\left(\boldsymbol{x}+\boldsymbol{e}_\alpha\delta t, t+\delta t\right)=\left(1-\frac{1}{\tau}\right)f_\alpha\left(\boldsymbol{x},t\right)+\frac{1}{\tau}f_\alpha^{eq}\left(\boldsymbol{x},t\right), \tag{8}$$

where $\tau$, $\delta t$, $f_\alpha$, and $\boldsymbol{e}_\alpha$ are the single relaxation time, time step, particle distribution function (PDF) along the $\alpha$ direction at lattice site $\boldsymbol{x}$, and discrete particle velocity vector in the $\alpha$ direction, respectively. $f_\alpha^{eq}$ is the local EDF, which can be written as

$$f_\alpha^{eq}\left(\boldsymbol{x},t\right)=w_\alpha\rho\left(1+\frac{\boldsymbol{e}_\alpha\cdot\boldsymbol{u}}{c_s^2}+\frac{\left(\boldsymbol{e}_\alpha\cdot\boldsymbol{u}\right)^2}{2c_s^4}-\frac{\boldsymbol{u}\cdot\boldsymbol{u}}{2c_s^2}\right). \tag{9}$$

where $\rho$, $\boldsymbol{u}$, $c_s$, and $w_\alpha$ are the macroscopic density, macroscopic velocity vector, lattice sound speed, and weighting coefficient, respectively.

The standard LBM generally consists of three steps, where the first step is the collision operation described by

$$f_\alpha^*(\boldsymbol{x},t) = \left(1 - \frac{1}{\tau}\right) f_\alpha(\boldsymbol{x},t) + \frac{1}{\tau} f_\alpha^{eq}(\boldsymbol{x},t), \tag{10}$$

where $f_\alpha^*$ is the post-collision state. The second step is the streaming propagation operation

$$f_\alpha(\boldsymbol{x}, t + \delta t) = f_\alpha^*(\boldsymbol{x} - e_\alpha \delta t, t), \tag{11}$$

which means that the post-collision state $f_\alpha^*$ propagates along direction $\alpha$. The final step involves computing the macroscopic flow variables: density ($\rho$) and momentum ($\rho \boldsymbol{u}$), which are yielded from the zeroth and first moments of the PDF,

$$\rho = \sum f_\alpha, \ \rho \boldsymbol{u} = \sum e_\alpha f_\alpha. \tag{12}$$

Through Chapman-Enskog (C-E) expansion analysis (see Section SI-3 in the Supplementary Information), it can be shown that the LBE recovers the following weakly compressible Navier-Stokes (N-S) equations with second-order accuracy in both space and time [41],

$$\begin{aligned} \frac{\partial \rho}{\partial t} + \nabla \cdot (\rho \boldsymbol{u}) &= 0, \\ \frac{\partial \rho \boldsymbol{u}}{\partial t} + \nabla \cdot (\rho \boldsymbol{u} \boldsymbol{u}) &= -\nabla p + \nabla \cdot \left[ \mu \left( \nabla \boldsymbol{u} + (\nabla \boldsymbol{u})^T \right) \right]. \end{aligned} \tag{13}$$

Here, $p$ and $\mu$ are the pressure and dynamic viscosity, respectively. The dynamic viscosity $\mu$ is related to the single relaxation time $\tau$ by

$$\mu = \rho c_s^2 \left( \tau - \frac{1}{2} \right) \delta t. \tag{14}$$

The LBM offers several advantages, including simple algebraic operations, ease of implementation, and efficient parallelization. However, it requires significantly more virtual memory to store the PDF compared to traditional N-S solvers [27]. To reduce memory requirements, Inamuro [39] introduced a lattice kinetic scheme where the relaxation time is fixed at $\tau = 1$, simplifying the LBE to:

$$f_\alpha\left(\boldsymbol{x} + \boldsymbol{e}_\alpha \delta t, t + \delta t\right) = f_\alpha^{eq}\left(\boldsymbol{x}, t\right), \tag{15}$$

Consequently, the collision and streaming steps become

$$\begin{aligned} f_\alpha^*\left(\boldsymbol{x}, t\right) &= f_\alpha^{eq}\left(\boldsymbol{x}, t\right), \\ f_\alpha\left(\boldsymbol{x}, t + \Delta t\right) &= f_\alpha^*\left(\boldsymbol{x} - e_\alpha \delta t, t\right), \end{aligned} \tag{16}$$

and the macroscopic quantities can be directly computed from the EDF $f_\alpha^{eq}$ by

$$\begin{aligned} \rho\left(\boldsymbol{x}, t + \delta t\right) &= \sum f_\alpha^{eq}\left(\boldsymbol{x} - e_\alpha \delta t, t\right), \\ \rho\left(\boldsymbol{x}, t + \delta t\right)\boldsymbol{u}\left(\boldsymbol{x}, t + \delta t\right) &= \sum e_\alpha f_\alpha^{eq}\left(\boldsymbol{x} - e_\alpha \delta t, t\right). \end{aligned} \tag{17}$$

Since the right-hand side of Eq. (15) consists solely of the EDF $f_\alpha^{eq}$, storing the full PDF becomes unnecessary, reducing memory demands. However, when using the standard EDF form as Eq. (9), the viscosity $\mu$ is fixed as

$$\mu = \frac{1}{2}\rho \mathrm{c}_s^2 \delta t = \rho \frac{\left(\Delta x\right)^2}{6\delta t}. \tag{18}$$

In LBM, the time step $\delta t$ and the spatial mesh spacing $\Delta x$ are related by $\Delta x = c \cdot \delta t$, where $c = 1$ represents the dimensionless particle velocity. Applying this relationship to Eq. (18), we can conclude that for a fixed mesh size, the classic LKS can only simulate flows with a specific viscosity. Consequently, simulating flows at different Reynolds numbers requires adjusting the mesh size. For high Reynolds number flows, this leads to extremely fine mesh spacing, significantly increasing the number of mesh points and the overall computational cost. To overcome this limitation, Inamuro [39] proposed a modification to the EDF as follows

$$f_\alpha^{eq}(\boldsymbol{x},t) = w_\alpha \rho \left( 1 + \frac{e_\alpha \cdot \boldsymbol{u}}{c_s^2} + \frac{(e_\alpha \cdot \boldsymbol{u})^2}{2c_s^4} - \frac{\boldsymbol{u} \cdot \boldsymbol{u}}{2c_s^2} + A \delta t e_\alpha^T \left( \nabla \boldsymbol{u} + (\nabla \boldsymbol{u})^T \right) e_\alpha \right). \qquad (19)$$

Through the C-E expansion analysis (see Appendix A), the viscosity $\mu$ is then related to the constant $A$ by

$$\mu = \rho \left( \frac{1}{2} - 2A c_s^2 \right) c_s^2 \delta t. \qquad (20)$$

It can be seen from Eq. (20) that for a given Reynolds number, the viscosity $\mu$ can be flexibly adjusted via the constant $A$. The introduction of constant $A$ removes the constraint on mesh size selection while retaining the simplification $\tau = 1$.

## Data availability

All data needed to evaluate the conclusions in the paper will openly available on GitHub at https://github.com/XiaoY-1012/QLKS-LBM upon publication.

## Code availability

All the source codes to reproduce the results in this study will be openly available on GitHub at https://github.com/XiaoY-1012/QLKS-LBM upon publication.

## Acknowledgments

The research is partially supported by the National Natural Science Foundation of China (92271103, 12202191) and the Research Fund of State Key Laboratory of Mechanics and Control for Aerospace Structures (Nanjing University of Aeronautics and Astronautics) (MCAS-I-0324G04).

**Author contributions**

L.M.Y. and C.S. contributed to the ideation and design of the research; Y.X. performed the research (implemented the model, conducted numerical experiments, analyzed the data); Y.X., Y.J.D., H.D. and J.W. provided CFD theoretical support and numerical benchmarks; Y.X. wrote the manuscript; L.M.Y. and C.S. contributed to manuscript editing; All authors participated in discussions and contributed with ideas.

**Competing interests**

The authors declare no competing interests.

# Supplementary Information for "Quantum Lattice Kinetic Scheme for Solving Two-dimensional and Three-dimensional Incompressible Flows"

Yang Xiao[1], Liming Yang[1, 2, 3, *], Chang Shu[4, †], Yinjie Du[1], Hao Dong[1], Jie Wu[1]

[1]Department of Aerodynamics, College of Aerospace Engineering, Nanjing University of Aeronautics and Astronautics, Nanjing 210016, China

[2]State Key Laboratory of Mechanics and Control for Aerospace Structures, Nanjing University of Aeronautics and Astronautics, Nanjing 210016, China

[3]MIIT Key Laboratory of Unsteady Aerodynamics and Flow Control, Nanjing University of Aeronautics and Astronautics, Nanjing 210016, China

[4]Department of Mechanical Engineering, National University of Singapore, Singapore 117576, Singapore

In this Supplementary Information, we have further supplemented the details of the paper.

## Table of Contents

**SI-1 Detailed supplementation of quantum lattice kinetic scheme**

The quantum circuit used for this initialization process in the D2Q9 lattice model is shown in Fig. S1.

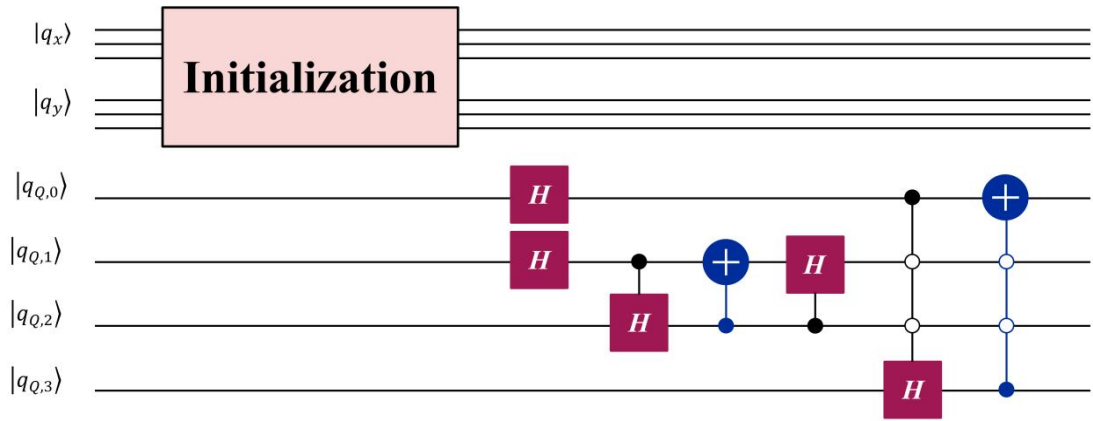

**Fig. S1** Quantum circuit for the initialization process in the D2Q9 lattice model.

The quantum circuit for computing the macroscopic density consists of a series of SWAP and Hadamard gates, as shown in Fig. S2. The complexity of the quantum circuit depends solely on the number of qubits in the register $|q_Q\rangle$, meaning it scales with the size of the velocity set rather than the mesh size.

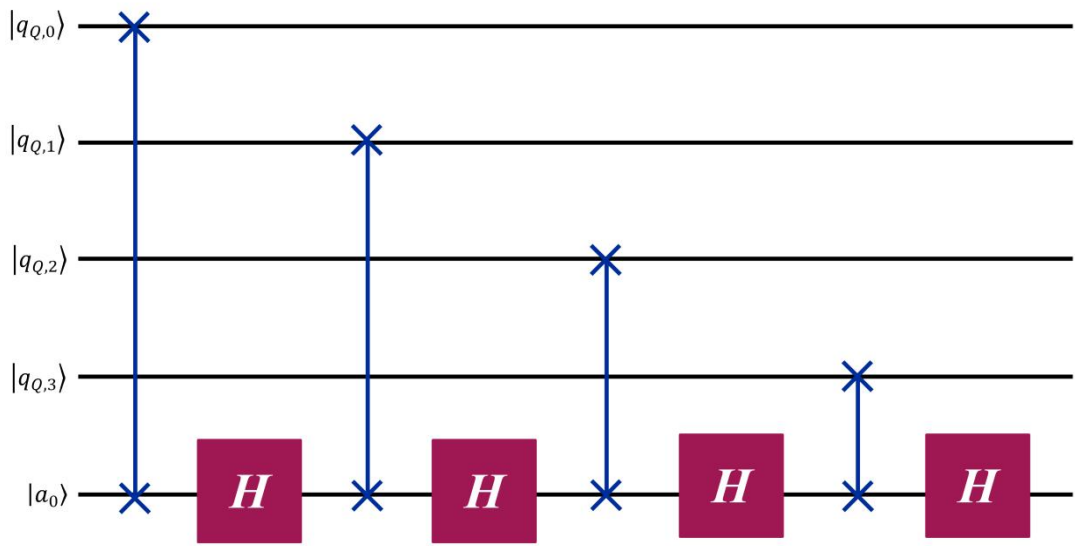

**Fig. S2** Quantum circuits for computing the macroscopic density in the D2Q9 lattice model.

## SI-2 The analytical solutions of 2D and 3D Taylor-Green vortex flows

The analytical solution of 2D Taylor-Green vortex flow is given by

$$\begin{cases} u(x,y,t) = -u_0 \cos\left(\dfrac{\pi x}{L}\right) \sin\left(\dfrac{\pi y}{L}\right) e^{\frac{-2\pi^2 u_0 t}{ReL}}, \\[2ex] v(x,y,t) = u_0 \sin\left(\dfrac{\pi x}{L}\right) \cos\left(\dfrac{\pi y}{L}\right) e^{\frac{-2\pi^2 u_0 t}{ReL}}, \\[2ex] \rho(x,y,t) = \rho_0 - \dfrac{\rho_0 u_0^2}{4c_s^2}\left[\cos\left(\dfrac{2\pi x}{L}\right) + \cos\left(\dfrac{2\pi y}{L}\right)\right] e^{\frac{-4\pi^2 u_0 t}{ReL}}. \end{cases} \tag{SI-2.1}$$

The analytical solution of 3D Taylor-Green vortex flow is given by

$$\begin{cases} u(x,y,z,t) = -u_0 \cos\left(\dfrac{\pi x}{L}\right) \sin\left(\dfrac{\pi y}{L}\right) e^{\frac{-2\pi^2 u_0 t}{ReL}}, \\[2ex] v(x,y,z,t) = u_0 \sin\left(\dfrac{\pi x}{L}\right) \cos\left(\dfrac{\pi y}{L}\right) e^{\frac{-2\pi^2 u_0 t}{ReL}}, \\[2ex] w(x,y,z,t) = 0, \\[2ex] \rho(x,y,z,t) = \rho_0 - \dfrac{\rho_0 u_0^2}{4c_s^2}\left[\cos\left(\dfrac{2\pi x}{L}\right) + \cos\left(\dfrac{2\pi y}{L}\right)\right] e^{\frac{-4\pi^2 u_0 t}{ReL}}. \end{cases} \tag{SI-2.2}$$

Here, $u_0$, $\rho_0$, $c_s$, and $Re$ are the characteristic velocity, reference density, sound speed, and Reynolds number, respectively.

## SI-3 Chapman–Enskog analysis for lattice kinetic scheme

We begin by rewriting Eq. (8) with the EDF defined as Eq. (19) into the following form with a source term

$$f_\alpha(\boldsymbol{x} + \boldsymbol{e}_\alpha \delta t, t + \delta t) - f_\alpha(\boldsymbol{x}, t) = -\frac{1}{\tau}\left(f_\alpha(\boldsymbol{x}, t) - f_\alpha^{eq}(\boldsymbol{x}, t)\right) + \delta t F_\alpha, \tag{SI-3.1}$$

where

$$f_\alpha^{eq}(\boldsymbol{x},t) = w_\alpha \rho \left( 1 + \frac{e_\alpha \cdot \boldsymbol{u}}{c_s^2} + \frac{(e_\alpha \cdot \boldsymbol{u})^2}{2c_s^4} - \frac{\boldsymbol{u} \cdot \boldsymbol{u}}{2c_s^2} \right),$$

$$F_\alpha = \frac{w_\alpha \rho A e_\alpha^T \left( \nabla \boldsymbol{u} + (\nabla \boldsymbol{u})^T \right) e_\alpha}{\tau}. \tag{SI-3.2}$$

Applying a second-order Taylor expansion around $(\boldsymbol{x}, t)$ to the left-hand side of Eq. (SI-3.1), we obtain

$$\delta t \left( \frac{\partial}{\partial t} + e_\alpha \cdot \nabla \right) f_\alpha + \frac{\delta t^2}{2} \left( \frac{\partial}{\partial t} + e_\alpha \cdot \nabla \right)^2 f_\alpha + \frac{1}{\tau} \left( f_\alpha - f_\alpha^{eq} \right) - \delta t F_\alpha = 0, \tag{SI-3.3}$$

Then, a series of Chapman-Enskog (C-E) expansions is introduced as follows

$$f_\alpha = f_\alpha^{(0)} + \varepsilon f_\alpha^{(1)} + \varepsilon^2 f_\alpha^{(2)} + \cdots,$$

$$\frac{\partial}{\partial t} = \varepsilon \frac{\partial}{\partial t_1} + \varepsilon^2 \frac{\partial}{\partial t_2} + \cdots,$$

$$\frac{\partial}{\partial \boldsymbol{x}} = \varepsilon \frac{\partial}{\partial \boldsymbol{x}_1} + \cdots, \tag{SI-3.4}$$

$$F_\alpha = \varepsilon F_\alpha^{(1)} + \cdots,$$

where $\varepsilon$ denotes the expansion parameter associated with the ratio between the lattice unit and the characteristic length scale [S1].

By substituting Eq. (SI-3.4) into Eq. (SI-3.3) and collecting terms at each order of $\varepsilon$, we obtain

$$O(\varepsilon^0): \quad \frac{1}{\tau \delta t} \left( f_\alpha^{(0)} - f_\alpha^{eq} \right) = 0, \tag{SI-3.5}$$

$$O(\varepsilon^1): \quad \frac{1}{\tau \delta t} f_\alpha^{(1)} + D_1 f_\alpha^{(0)} - F_\alpha^{(1)} = 0, \tag{SI-3.6}$$

$$O(\varepsilon^2): \quad \frac{1}{\tau \delta t} f_\alpha^{(2)} + D_1 f_\alpha^{(1)} + \frac{\partial f_\alpha^{(0)}}{\partial t_2}$$

$$+ \frac{\delta t}{2} \left( \frac{\partial^2 f_\alpha^{(0)}}{\partial t_1^2} + 2 e_\alpha \cdot \frac{\partial^2 f_\alpha^{(0)}}{\partial t_1 \partial \boldsymbol{x}_1} + e_\alpha e_\alpha : \frac{\partial^2 f_\alpha^{(0)}}{\partial \boldsymbol{x}_1^2} \right) = 0, \tag{SI-3.7}$$

where $D_k = \frac{\partial}{\partial t_k} + e_\alpha \cdot \frac{\partial}{\partial \boldsymbol{x}_k}$. By applying Eq. (AI-3.6), Eq. (AI-3.7) can be further simplified as follows

$$O\left(\varepsilon^2\right): \quad \frac{1}{\tau\delta t}f_\alpha^{(2)} + \left(1 - \frac{1}{2\tau}\right)D_1 f_\alpha^{(1)} + \frac{\partial f_\alpha^{(0)}}{\partial t_2} + \frac{\delta t}{2}D_1 F_\alpha^{(1)} = 0. \quad \text{(SI-3.8)}$$

Next, we recover the macroscopic equations by calculating the velocity moments in Eqs. (SI-3.5)–(SI-3.8). Based on Eqs. (SI-3.2) and (SI-3.5), the distribution function $f_\alpha^{(0)}$ satisfies

$$\begin{cases} f_\alpha^{(0)} = f_\alpha^{eq}, \\ \sum_\alpha f_\alpha^{(0)} = \sum_\alpha f_\alpha^{eq} = \rho, \\ \sum_\alpha e_\alpha f_\alpha^{(0)} = \sum_\alpha e_\alpha f_\alpha^{eq} = \rho\boldsymbol{u}, \\ \sum_\alpha e_\alpha e_\alpha f_\alpha^{(0)} = \sum_\alpha e_\alpha e_\alpha f_\alpha^{eq} = \rho\boldsymbol{u}\boldsymbol{u} + \rho c_s^2 \boldsymbol{I}. \end{cases} \quad \text{(SI-3.9)}$$

Combining Eq. (SI-3.9) with the truncated expansion $f_\alpha = f_\alpha^{(0)} + \varepsilon f_\alpha^{(1)} + \varepsilon^2 f_\alpha^{(2)}$ of Eq. (SI-3.4) and according to the compatibility condition, we obtain

$$\begin{cases} \sum_\alpha f_\alpha^{(1)} = \sum_\alpha f_\alpha^{(2)} = 0, \\ \sum_\alpha e_\alpha f_\alpha^{(1)} = 0, \end{cases} \quad \text{(SI-3.10)}$$

For the source term $F_\alpha$, it can be easily computed from Eq. (SI-3.2) as

$$\begin{cases} \sum_\alpha F_\alpha = 0, \\ \sum_\alpha e_\alpha F_\alpha = 0, \\ \sum_\alpha e_\alpha e_\alpha F_\alpha = 2\rho A c_s^4 \left[\nabla\boldsymbol{u} + (\nabla\boldsymbol{u})^T\right] \big/ \tau. \end{cases} \quad \text{(SI-3.11)}$$

Based on Eqs. (SI-3.9)-(SI-3.11), the 0th-order and 1st-order velocity moments of Eq. (SI-3.6) can be computed as

$$\frac{\partial \rho}{\partial t_1} + \nabla_1 \cdot (\rho\boldsymbol{u}) = 0, \quad \text{(SI-3.12)}$$

$$\frac{\partial(\rho\boldsymbol{u})}{\partial t_1} + \nabla_1 \cdot \left(\rho\boldsymbol{u}\boldsymbol{u} + \rho c_s^2 \boldsymbol{I}\right) = 0. \quad \text{(SI-3.13)}$$

Similarly, the 0th-order and 1st-order velocity moments of Eq. (SI-3.8) are

$$\frac{\partial \rho}{\partial t_2} = 0, \tag{SI-3.14}$$

$$\frac{\partial (\rho \boldsymbol{u})}{\partial t_2} + \left(1 - \frac{1}{2\tau}\right) \nabla_1 \cdot \left(\sum_\alpha e_\alpha e_\alpha f_\alpha^{(1)}\right) = -\frac{\delta t}{2} \nabla_1 \cdot \left(\sum_\alpha e_\alpha e_\alpha F_\alpha^{(1)}\right). \tag{SI-3.15}$$

From Eq. (SI-3.6), $f_\alpha^{(1)}$ can be expressed as

$$f_\alpha^{(1)} = -\tau \delta t \left(D_1 f_\alpha^{(0)} - F_\alpha^{(1)}\right) \tag{SI-3.16}$$

Therefore, $\sum_\alpha e_\alpha e_\alpha f_\alpha^{(1)}$ becomes

$$
\begin{aligned}
\sum_\alpha e_\alpha e_\alpha f_\alpha^{(1)} &= -\tau \delta t \left[ D_1 \left(\sum_\alpha e_\alpha e_\alpha f_\alpha^{(0)}\right) - \sum_\alpha e_\alpha e_\alpha F_\alpha^{(1)} \right] \\
&= -\tau \delta t \left[ \frac{\partial (\rho \boldsymbol{uu})}{\partial t_1} + c_s^2 \frac{\partial \rho}{\partial t_1} I + \nabla_1 \cdot \left(\sum_\alpha e_\alpha e_\alpha e_\alpha f_\alpha^{(0)}\right) - \sum_\alpha e_\alpha e_\alpha F_\alpha^{(1)} \right].
\end{aligned} \tag{SI-3.17}
$$

From Eq. (SI-3.12), we further derive

$$\frac{\partial}{\partial t_1}\left(\rho u_i u_j + \rho c_s^2 \delta_{ij}\right) = \frac{\partial (\rho u_i u_j)}{\partial t_1} - c_s^2 \nabla_1 \cdot (\rho \boldsymbol{u}) \delta_{ij}. \tag{SI-3.18}$$

The term $\partial \rho u_i u_j / \partial t_1$ can be rewritten as

$$\frac{\partial \rho u_i u_j}{\partial t_1} = u_j \frac{\partial (\rho u_i)}{\partial t_1} + u_i \frac{\partial (\rho u_j)}{\partial t_1} - u_i u_j \frac{\partial \rho}{\partial t_1}. \tag{SI-3.19}$$

Based on Eqs. (SI-3.12) and (SI-3.13), we obtain

$$\frac{\partial \rho u_i u_j}{\partial t_1} = -c_s^2 u_j \frac{\partial \rho}{\partial x_{1i}} - c_s^2 u_i \frac{\partial \rho}{\partial x_{1j}} - \frac{\partial}{\partial x_{1k}}\left(\rho u_i u_j u_k\right). \tag{SI-3.20}$$

By performing a simple algebraic calculation based on Eq. (SI-3.2), we obtain

$$
\begin{aligned}
\frac{\partial}{\partial x_{1k}}\left(\sum_\alpha e_{\alpha i} e_{\alpha j} e_{\alpha k} f_\alpha^{(0)}\right) &= \frac{\partial}{\partial x_{1k}}\left[\rho c_s^2\left(\delta_{ij} u_k + \delta_{ik} u_j + \delta_{jk} u_i\right)\right] \\
&= c_s^2 \nabla_1 \cdot (\rho \boldsymbol{u}) \delta_{ij} + \rho c_s^2 \frac{\partial u_j}{\partial x_{1i}} + c_s^2 u_j \frac{\partial \rho}{\partial x_{1i}} + \rho c_s^2 \frac{\partial u_i}{\partial x_{1j}} + c_s^2 u_i \frac{\partial \rho}{\partial x_{1j}}.
\end{aligned} \tag{SI-3.21}
$$

By simultaneously applying Eqs. (SI-3.18)-(SI-3.21), we obtain

$$\sum_\alpha e_\alpha e_\alpha f_\alpha^{(1)} = -\tau \delta t \left[ \rho c_s^2 \left( \nabla \boldsymbol{u} + (\nabla \boldsymbol{u})^T \right) - \frac{\partial}{\partial \boldsymbol{x}_{1k}} \left( \rho u_i u_j u_k \right) - \sum_\alpha e_\alpha e_\alpha F_\alpha^{(1)} \right]. \quad \text{(SI-3.22)}$$

Neglecting the higher-order terms $\dfrac{\partial}{\partial \boldsymbol{x}_{1k}} \left( \rho u_i u_j u_k \right)$ yields

$$\frac{\partial (\rho \boldsymbol{u})}{\partial t_2} + \left( 1 - \frac{1}{2\tau} \right) \nabla_1 \cdot \rho \left[ \nabla \boldsymbol{u} + (\nabla \boldsymbol{u})^T \right] = -\delta t \, \tau \nabla_1 \cdot \left( \sum_\alpha e_\alpha e_\alpha F_\alpha^{(1)} \right). \quad \text{(SI-3.23)}$$

By applying the reverse multi-scale expansion to Eqs. (SI-3.12)-(SI-3.14) and (SI-3.23), the macroscopic equations in the original scale can be obtained. Specifically, by applying $\varepsilon \times$ Eq. (SI-3.12) + $\varepsilon^2 \times$ Eq. (SI-3.14) yields

$$\frac{\partial \rho}{\partial t} + \nabla \cdot (\rho \boldsymbol{u}) = 0. \quad \text{(SI-3.24)}$$

Similarly, combining Eqs. (SI-3.13), (SI-3.23), and (SI-3.11) yields

$$\frac{\partial (\rho \boldsymbol{u})}{\partial t} + \nabla \cdot (\rho \boldsymbol{u} \boldsymbol{u}) = -\nabla \left( \rho c_s^2 \right) + \delta t c_s^2 \left( \tau - \frac{1}{2} - 2 A c_s^2 \right) \nabla \cdot \rho \left[ \nabla \boldsymbol{u} + (\nabla \boldsymbol{u})^T \right] \quad \text{(SI-3.25)}$$

Comparing Eqs. (SI-3.24)–(SI-3.25) with Eq. (13), it can be seen that in LBM, the fluid pressure is defined as

$$p = \rho c_s^2, \quad \text{(SI-3.26)}$$

and the fluid viscosity is defined as

$$\mu = \rho \left( \tau - \frac{1}{2} - 2 A c_s^2 \right) \delta t c_s^2. \quad \text{(SI-3.27)}$$

From Eq. (SI-3.27), when $A = 0$, it reduces to the viscosity formula (Eq. (14)) of the classic LBM. When $A = 0$ and $\tau = 1$, it reduces to the viscosity formula (Eq. (18)) of the classic LKS.